\documentclass[12pt,superscriptaddress,showpacs,showkeys,twocolumn,prb,aps,reprint,a4paper,floatfix]{revtex4-2}

\usepackage{graphicx}
\usepackage{amsmath, amsthm, amssymb, amsfonts}
\usepackage{float}
\usepackage{natbib}
\usepackage{physics}
\usepackage[para,online,flushleft]{threeparttable} %For tablenotes
\usepackage{array}
\usepackage{ragged2e} %For left justification

\makeatletter
\newcommand*{\citenst}[2][]{%
  \begingroup
  \let\NAT@mbox=\mbox
  \let\@cite\NAT@citenum
  \let\NAT@space\NAT@spacechar
  \let\NAT@super@kern\relax
  \renewcommand\NAT@open{[}%
  \renewcommand\NAT@close{]}%
  \citet[#1]{#2}%
  \endgroup
}
\makeatother

\makeatletter
\newcommand*{\citenumns}[2][]{%
  \begingroup
  \let\NAT@mbox=\mbox
  \let\@cite\NAT@citenum
  \let\NAT@space\NAT@spacechar
  \let\NAT@super@kern\relax
  \renewcommand\NAT@open{[}% What bracket you wish to use
  \renewcommand\NAT@close{]}%
  \cite[#1]{#2}% Here is the difference!
  \endgroup
}
\makeatother
\usepackage[caption=false]{subfig}

\begin{document}
\title{Superconducting resonator parametric amplifiers with intrinsic separation of pump and signal tones}
\author{Songyuan Zhao}
\email{Author to whom any correspondence should be addressed.\\ E-mail: sz311@cam.ac.uk}
\author{S. Withington}
\author{C. N. Thomas}
\date{\today}

\affiliation{Cavendish Laboratory, JJ Thomson Avenue, Cambridge CB3 OHE, United Kingdom.}

\begin{abstract}
\noindent Superconducting resonator parametric amplifiers achieve ultra-low-noise amplification through the nonlinear kinetic inductance of thin-film superconductors. One of the main challenges to the operation of these devices is the separation of the strong pump tone from the signal tone after amplification has been achieved. In this paper, we propose and experimentally demonstrate a pump separation method based on operating a half-wave superconducting resonator amplifier behind a cryogenic circulator. Our pump separation scheme does not involve post-amplification interference, and thereby avoids the delicate phase matching of two different pump paths. We demonstrate the scheme using two-port half-wave resonator amplifiers based on superconducting NbN thin-films. We present measurements of gain profiles and degrees of pump separation for amplifiers having different coupling quality factors. On an amplifier having a coupling quality factor of $\sim2000$, we measured a peak signal gain of $15\,\mathrm{dB}$ whilst achieving pump separation of $12\,\mathrm{dB}$. The amplifier was stable for continuous measurements, and the gain drift was measured to be $0.15\,\mathrm{dB}$ over an hour. The same amplifier was operated at $3.2\,\mathrm{K}$ and achieved a peak signal gain of $11\,\mathrm{dB}$ whilst having a pump separation factor of $10.5\,\mathrm{dB}$. The pump separation scheme, and these promising results, will advance the development of superconducting resonator amplifiers as an important technology in quantum sensing.
\end{abstract}


\maketitle

\section{Introduction}

Superconducting amplifiers are ultra-low-noise electronic devices that achieve amplification by exploiting the nonlinear kinetic inductance of thin-film superconductors \citenumns{Eom_2012,zhao2022physics}. These devices have the potential to significantly improve the performance of a wide variety of fundamental physics experiments and utilitarian technological applications, particularly in areas such as quantum computing \citenumns{Ranzani_2018,Zobrist_2019,Vissers_2020, Project8_2009, Saakyan_2020}. In this paper, we propose and demonstrate an elegant and cost-effective way of operating superconducting resonator parametric amplifiers that addresses the essential need to separate the strong pump tone from the signal tone once amplification has taken place. If this is not done, the transmitted pump tone can limit dynamic range, cause later stages of the amplifier chain to saturate, and generate spurious harmonics that are problematic for system fidelity and stability.

% General introduction to parametric amplifiers - TWPAs and resonator amplifiers
Signal amplification in superconducting parametric amplifiers is achieved through wave-mixing processes induced by the underlying nonlinearity, both reactive and dissipative, of the superconducting thin-film devices\citenumns{Eom_2012,Zhao_2022,Zhao_2023}. 
At present, there are two main types of superconducting parametric amplifier technologies: Josephson Parametric Amplifiers which rely on the nonlinear Josephson junction inductance \citenumns{Aumentado_JPA_2020,Yurke_1988,Movshovich_1990}, and Kinetic Inductance Parametric Amplifiers which rely on the nonlinear kinetic inductance of superconducting thin-films \citenumns{Eom_2012,Jonas_review}. In this study, we focus on the latter. Experimentally, kinetic inductance amplifiers have demonstrated gains of $10-30 \,\mathrm{dB}$ whilst achieving noise performance close to the standard quantum limit for linear amplifiers of half a quantum \citenumns{Tholen_2009, Eom_2012, Vissers_2016, Malnou_2020}. By operating in degenerate mixing mode, these amplifiers have also been used to realise squeezed amplification and generate squeezed vacuum states \citenumns{Parker_squeezing_2022}. Superconducting parametric amplifiers have been realised in both travelling-wave and resonator geometries. In general, amplifiers based on travelling-wave designs have wider bandwidths, whereas amplifiers based on resonator designs have lower pump power requirements and are less prone to lithographic defects such as shorts or breaks on long ($>0.5\,\mathrm{m}$) transmission lines with micron-scale features \citenumns{Eom_2012,Shan_2016,zhao2022physics}. Regardless of geometry, central to the process of parametric amplification is the use of a strong pump tone: the nonlinear wave-mixing processes typically transfer energy from the strong pump tone to the weak signal tone in order to achieve amplification. It is important that this strong pump tone (as large as $-8.0\,\mathrm{dBm}$ at the input of the amplifiers in some instances \citenumns{Eom_2012}) is separated from the amplified signal tone so that no unwanted saturation or nonlinearity occurs in the downstream electronics, especially cryogenic high electron mobility transistor (HEMT) amplifiers. Further, in the case of four-wave-mixing, the pump frequency is located close to the peak of the signal gain profile. As a result, this region of maximum gain may be unusable due to contamination of the signal by the phase-noise of the pump.

% Interference based pump separation scheme
In our work to date, the pump tone has been separated off from the post-amplified signal by using an interference scheme: 1) Before entering the cryostat, the pump is split into two paths; 2) one of the paths is sent to the parametric amplifier as the pump and the other is maintained as a reference; 3) after amplification, the two paths are recombined out-of-phase to remove the pump tone through destructive interference. Whilst this scheme is effective in attenuating the pump tone by $\sim20\,\mathrm{dB}$ \citenumns{zhao2022physics}, there are a number of practical constraints: 1) It is difficult to match the power in the two paths to the required degree. A continuous analogue variable attenuator can be used, but it requires manually adjusting by the experimenter. On the other hand, a programmable variable attenuator has discrete steps and this places a limit on the degree of matching. For $-8\,\mathrm{dBm}$ of pump power and a typical step size of $0.5\,\mathrm{dB}$ \citenumns{VarAtt}, the degree of separation is limited to $\sim10\,\mathrm{dB}$. 2) An additional cable into the cryostat is needed to accommodate the reference signal. Although this manual scheme is acceptable for one-off amplifiers in the laboratory, it prevents turn-key operation, and it is not feasible when an array of amplifiers is needed because the additional electronics significantly increases the complexity of the cryostat and places additional heat load on the cold stage. 3) For resonator-based amplifiers, the pump is typically placed near resonance where the phase varies the most rapidly with frequency. This means that any change in properties of the resonator, due to say thermal fluctuations, results in a phase mismatch between the two paths and diminishes the effectiveness of interference. In addition, the stability of the amplifier is degraded by introducing this phase-to-amplitude convertor.

% Our scheme - exploit the asymmetry between pump and signal % Advantages
In this paper, we propose and demonstrate an elegant and cost-effective method of achieving pump-signal separation in superconducting parametric amplifiers based on two-port resonators. Our scheme does not rely on destructive interference, but instead exploits the asymmetry in behaviour between signal and pump through the use of a cryogenic circulator: in a two-port resonator parametric amplifier, the amplified signal appears in both the transmission mode as well as the reflection mode; the pump, on the other hand, can be strategically placed such that it primarily goes into the transmission mode. The amplified signal can thus be separated from the pump by using the reflection mode through a cryogenic circulator. This important characteristic will be discussed in more detail in the Theory section.

% Our measurements - depending on material
We have experimentally realised the pump separation scheme using two-port half-wave resonator amplifiers based on superconducting NbN. We first report a proof-of-concept experiment, which achieved pump separation of $5.5\,\mathrm{dB}$ whilst having a peak signal gain of $19\,\mathrm{dB}$. In this context, pump separation measures the net reduction in pump power in the output port relative to the input port, and signal gain measures the net amplification of signal power in the output port relative to the input port. We then demonstrate that the degree of pump separation can be enhanced when the quality factor of the resonator is limited by its coupling quality factor. In this set of experiments, we measured a pump separation of $12\,\mathrm{dB}$ whilst achieving a peak signal gain of $15\,\mathrm{dB}$. The amplifier was stable for continuous measurements, and the gain drift was measured to be $0.15\,\mathrm{dB}$ per hour. The same amplifier was also operated at $3.2\,\mathrm{K}$ where it achieved a peak signal gain of $11\,\mathrm{dB}$ whilst having a pump separation factor of $10.5\,\mathrm{dB}$. The degree of separation achieved compares favourably with the destructive interference scheme, and given the underlying nonlinear dynamics, higher degrees of pump separation are likely to be achievable by using a resonator with stronger coupling. Pump separation of $>10\,\mathrm{dB}$ is a highly meaningful result as it is already sufficient to prevent the saturation of down-stream HEMTs in typical parametric amplifier systems. In addition, we have found that the optimum operating point is theoretically predictable, easily found in practice, and highly resilient to drifts in the system. Overall, because of these qualities, we believe that it has the potential to enable the large scale deployment of superconducting resonator parametric amplifier technology, as for example, integrated amplifier arrays for quantum-noise limited phased-array antenna systems \citenumns{withington2024quantum, Esfahani_2023}.

\section{Theory}

% Fig of the scheme
\begin{figure}[!htb]
\includegraphics[width=8.6cm]{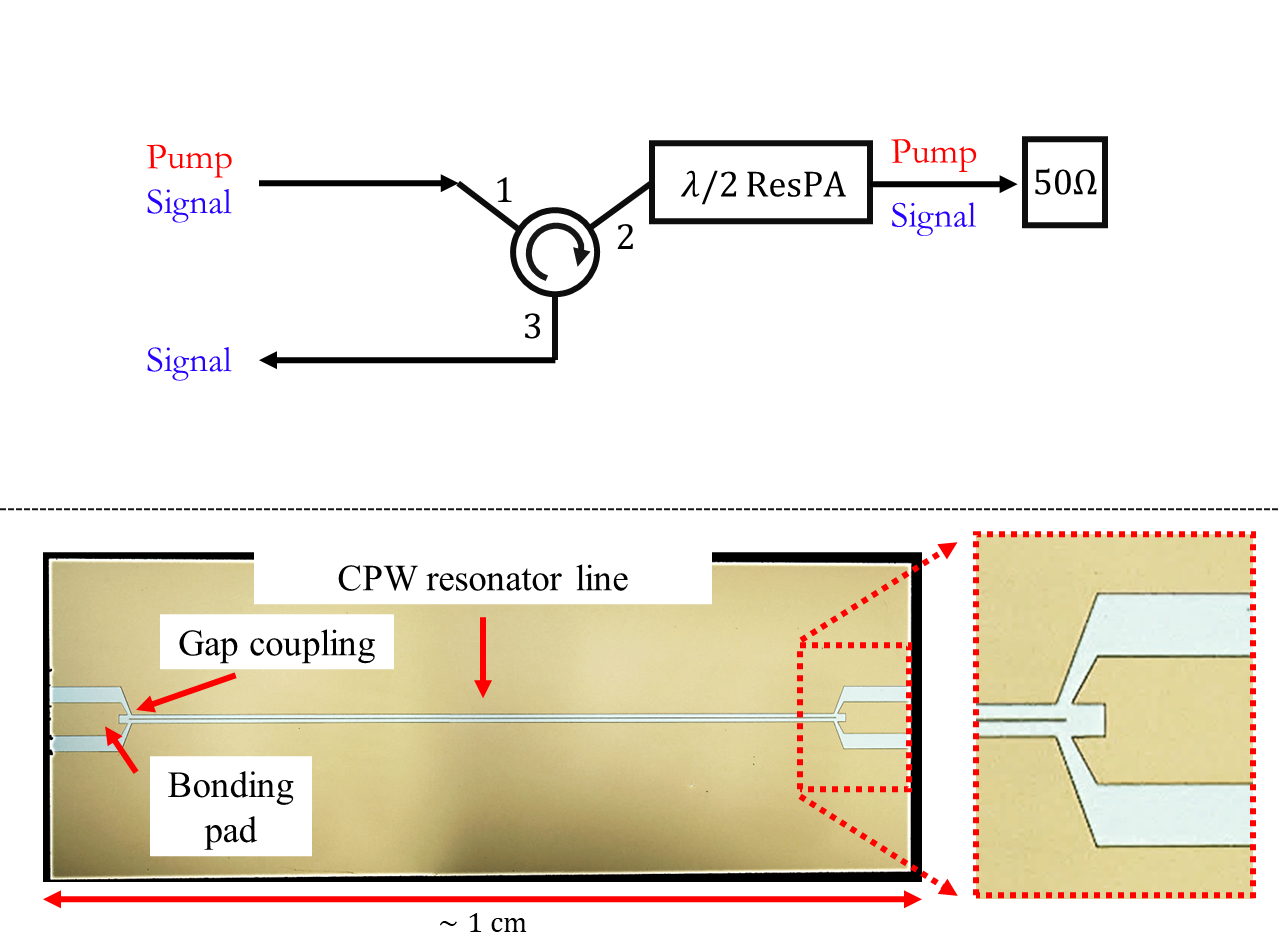}
\caption{\label{fig:schematic_diagram} Top subfigure: schematic diagram of the pump separation scheme. The pump tone and the signal tone are injected into port 1 of the circulator; the input tones are sent to the half-wave resonator amplifier which is connected to port 2 of the circulator; the transmission mode of the resonator amplifier, which contain both the amplified signal tone and the pump tone, are terminated at a matched $50\,\mathrm{\Omega}$ termination; the reflection mode of the resonator amplifier, which contains mostly the amplified signal tone, is received via port 3 of the circulator as output. \protect\linebreak Bottom subfigure: an example of the half-wave resonators measured in this study \citenumns{Zhao_2023}. The resonator comprises a length of coplanar waveguide capacitively coupled in series between two contact pads. The length over which the coupling pad overlaps with the resonator line determines the coupling quality factor of this resonator.}
\end{figure}

% Summary of principle

In previous experiments, we have successfully operated quarter-wavelength thin-film resonators as amplifiers in reflection mode, and half-wavelength thin-film resonators as amplifiers in transmission mode \citenumns{Zhao_2023}. In this study, we use half-wavelength resonators in reflection mode in order to exploit an asymmetry in the S-parameters of the pump and signal tones in two-port series-resonant structures. When a pump tone drives the input of a nonlinear series-resonant structure, it is mostly transmitted, and so in this sense the device is directional. When a signal is applied to the structure, however, approximately half of the amplified signal emerges from the output port, and the other half emerges from the input port. This occurs because although the amplifier has the form of a two-port network, it is essentially a negative-resistance device, and so amplified signal appears at both ports. This asymmetry allows the pump tone to be placed at a frequency that is strongly transmissive, and dissipated in a load, while the desired amplified signal can be collected via the reflection mode from the input port using a cryogenic circulator: figure~\ref{fig:schematic_diagram}.

\begin{figure}[!htb]
\includegraphics[width=8.6cm]{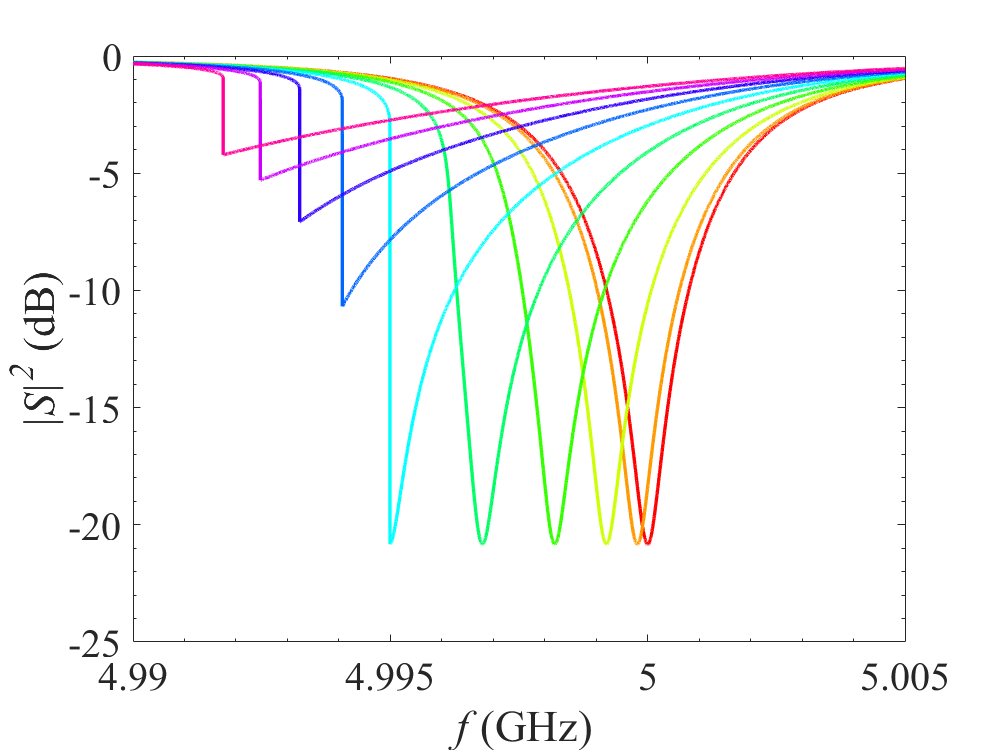}
\caption{\label{fig:0} Squared S-parameter of the reflection mode of a half-wave two port resonator at different sweep powers, simulated using the theory presented in \citenumns{Thomas_2022}. The sweep direction is from low frequency to high frequency, and the power increases from red to magenta. The curves are spaced linearly in the amplitude of the sweep tone such that the red curve at $5\,\mathrm{GHz}$ corresponds to negligibly small sweep power, and the cyan curve at $4.995\,\mathrm{GHz}$ corresponds to just enough power to produce a bifurcation discontinuity that is as deep as the resonance depth. Accordingly, the resonance response shows increasing distortion due to a reactive feedback process. Discontinuity in the resonance response develops when the sweep power is greater than that required to induce bifurcation.}
\end{figure}

% Theory summary
The nonlinear response of thin-film resonators as parametric amplifiers has been well-described by recent research and can be understood in two conceptual steps: The first step is to understand the nonlinear response of the resonator driven by a strong pump tone. This problem is relevant to other applications of superconducting resonators and is discussed in detail in \citenumns{Thomas_2020,Swenson_2013}, along with the physical origin of the nonlinearities. The second step is to understand the subsequent behaviour when additionally a weak signal is applied, as detailed in \citenumns{Thomas_2022}. Equivalently, we can view the problem as one of establishing the operating point for the device by the pump, followed by the small signal behaviour around that point. 

In what follows, we will apply the analysis presented in \citenumns{Thomas_2022} to the superconducting resonator amplifier described in this paper. This analysis assumes the resonator amplifier can be modelled by a set of temporal coupled-mode equations, and that the resonant frequency and the Q-factor of the mode vary with the square of the mode amplitude (i.e. the energy stored in the resonator). The result is a generalised Duffing-type model of nonlinear resonator which takes into account the effect of reactive, dissipative, and rate-limited nonlinearities in the wave-mixing process. Using the model, we have simulated the behaviour of a resonator with total quality factor $Q_\mathrm{r}$ of $1000$, coupling quality factor $Q_\mathrm{c}$ of $1100$, and resonance frequency of $5\,\mathrm{GHz}$, which is representative of the devices fabricated in our laboratory. We have also assumed that the nonlinearity is purely reactive, i.e. only the resonant frequency changes with power; this is the ideal case, and the aim of this section is to present qualitatively the characteristic behaviour pertaining to parametric amplification and pump-signal separation. Both figures in this section are modelling S-parameter of the reflection mode of the resonator (received at port 3 of the circulator configured according to figure~\ref{fig:schematic_diagram}).

% Resonator and bifurcation of pump tone
In the presence of a strong tone, the stored and dissipated energies modify the circuit parameters of the resonator, which in turn change the frequency of the underlying, not directly observed, resonance. This reactive process results in key features in the nonlinear response such as shifts in resonance frequencies, distortions in S-parameter profiles, and hysteresis and bifurcations: i.e. the resonator responds differently depending on whether the frequency is swept up or swept down. As shown in figure~\ref{fig:0}, as the sweep power is increased, from the red line to the magenta line, the resonance shifts and distorts largely as a result of reactive parametric feedback. When the applied power is greater than that required for bifurcation, a discontinuity appears in the observed transmission profile. As described by our recent work on resonator parametric amplifiers \citenumns{Thomas_2022}, the onset of this discontinuity is associated with the formation of operating points that can be used for high-gain, low-noise amplification. In practice, the optimum pump power is approximately the applied power at bifurcation, and the optimum pump frequency is approximately the frequency at which the discontinuity first appears.

\begin{figure}[!htb]
\includegraphics[width=8.6cm]{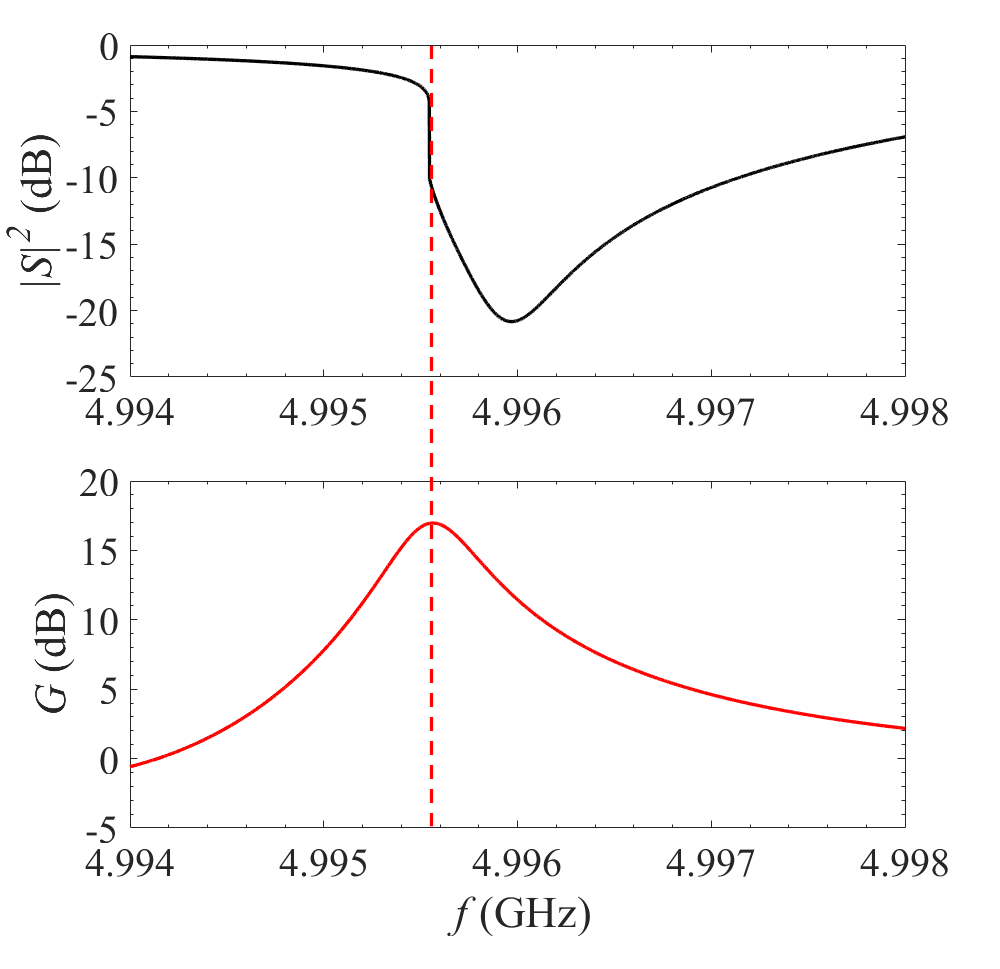}
\caption{\label{fig:1} Simulation of signal amplification and pump separation using the theory presented in \citenumns{Thomas_2022} at a fixed operating point. Top subfigure: resonance response of the reflection mode where the sweep power is equal to the pump power of the amplification operating point. Bottom subfigure: gain profile of the signal tone as a function of frequency. The pump tone is fixed at the operating point, and is represented by the dotted line. The dotted line intersects the top subfigure at frequency just above the discontinuity with a reflection coefficient of $-10.5\,\mathrm{dB}$. The peak gain of the bottom subfigure is $17\,\mathrm{dB}$.}
\end{figure}

% Amplification of signal tone
The parametric amplification of the signal tone is a result of the wave-mixing between the signal and the pump facilitated by the underlying nonlinearity. At high gain, in a resonator, this process generates amplified signals of similar magnitude propagating in both the transmission and the reflection directions. The top subfigure of figure~\ref{fig:1} shows the nonlinear resonance response in reflection where the power of the sweeping tone is equal to the pump power of the amplifier's operating point; the bottom subfigure shows the gain profile of the reflected signal tone as a function of frequency. The peak gain in the bottom subfigure is $17\,\mathrm{dB}$. The dashed line indicates the best  operating frequency of the pump tone; in the top subfigure, it intersects the pump response at a frequency just above the discontinuity with a reflection coefficient of $-10.5\,\mathrm{dB}$. This simulation illustrates that by operating a half-wavelength reflection amplifier with pump frequency just above the bifurcation frequency, the pump power reflected is significantly weaker than the pump power incident on the device; the signal tone, however, is strongly amplified in both the transmission mode and the reflection mode.

\section{Devices and measurements}
We have designed, fabricated, and packaged half-wave NbN resonators to experimentally realise the pump separation scheme. The resonators were fabricated in batches of related designs on $50.8\,\,\mathrm{mm}$ diameter, $225\,\,\mathrm{\mu m}$ thick, CZ silicon wafers. The NbN device layer was patterned via a lift-off process with image-reversal photoresist. These were then diced these into individual resonator chips after processing. The NbN films were deposited by reactive DC sputtering a niobium target in the presence of argon as the buffer gas and nitrogen as the reactive gas. All depositions were carried out in an ultra-high vacuum deposition system which has a base pressure of $2\times 10^{-10}\,\,\mathrm{Torr}$ or below. The measured superconducting transition temperatures of the films were $\sim10\,\mathrm{K}$. 

All devices tested in this study were half-wave resonators based on co-planar waveguide (CPW) geometries, as shown in the bottom subfigure of figure~\ref{fig:schematic_diagram}. The CPW conductor width was $5\,\mathrm{\mu m}$, and the gap width from conductor to ground was $50\,\mathrm{\mu m}$. These resonators were end coupled through capacitive small re-entrant lines. The film thicknesses were controlled by calibrating and varying the deposition time. All devices in this study had thicknesses of $\sim100\,\mathrm{nm}$. 

For testing, the resonators were bonded using aluminium wires to a gold-plated copper enclosure with SMA connectors. The enclosure was designed such that its resonance frequencies do not overlap significantly with the resonance frequencies of the half-wave resonators \citenumns{Tess_2021}. The packaged amplifier and the cryogenic circulator were arranged according to the top subfigure of figure~\ref{fig:schematic_diagram}. The cryogenic circulator used here was a QuinStar QCY series single-junction circulator with Mu-metal shield. Both the amplifier and the circulator were attached to the cold stage of an adiabatic demagnetization refrigerator. Stainless steel coaxial cables were used in the cryogenic system, thermally anchored at $35\,\mathrm{K}$, $4\,\mathrm{K}$, $1\,\mathrm{K}$, and $50\,\mathrm{mK}$ stages \citenumns{Zhao_2023}. HEMT amplifiers were not used in the signal chain since added-noise temperature measurement was not an aim of this study. The pump tone was generated and measured using a vector network analyser (VNA) and the signal tone was generated and measured using another VNA. The two VNAs were phase-locked to a common $10\,\mathrm{MHz}$ reference tone. The signal and the pump tone were combined at the input of the cryostat using a multiband two-way power combiner. In all measurements, both the pump and the signal tones were injected via port 1 and measured via port 3 of the circulator, where the ports were configured according to figure~\ref{fig:schematic_diagram}. Unless otherwise specified, all measurements were taken at cryogenic temperature of $0.1\,\mathrm{K}$, maintained using an adiabatic demagnetization refrigerator with a pulse tube cooler.

\subsection{Amplifier operating point}
% Method of finding operating point
Once the hardware components are configured properly, the experimenter still needs to find the operating point of the superconducting resonator parametric amplifier in terms of pump frequency and power to achieve both high gain and high degree of pump separation. Finding the operating point appears, at first sight, to be a complicated task. As a strong pump is swept across the notional resonance frequency, the energies stored and dissipated in the resonator change. Although the underlying resonance simply shifts and broadens, retaining a simple Lorentzian form, this is not what is observed by the experimenter in a swept measurement, which exhibits distortion and bifurcation. This effect is highly reproducible, and well described by our theory  \citenumns{Thomas_2022}, and so we have been able to establish a set-up procedure, which has proven to be exceedingly dependable and reproducible when applied to a wide range of resonator configurations and materials. The ruggedness of the operation is one of the key attractions of the resonator approach. The experimental procedure is as follows:

\begin{enumerate}
  \item Sweep the S-parameters of the resonator using a signal tone such that a clear, undistorted resonance can be observed.
  \item Increase the pump power until a discontinuity first appears in the frequency-swept S-parameters. Record the power of the pump tone and the frequency of discontinuity at the onset of bifurcation.
  \item Set the pump tone to have the recorded power and frequency. Sweep the S-parameters of the weak signal tone to verify amplification. The signal tone should be more than $40\,\mathrm{dB}$ weaker than the pump. If no amplification or only a small amplification is measured, adjust the pump frequency around the bifurcation frequency in small frequency steps, checking for amplification after each adjustment.
  \item Additionally, once a reliable operating point has been found, adjust the pump power and frequency around the operating point systematically to optimise for pump tone separation. In general, we have found that the optimum operating point is where there is a small discontinuity that does not overwhelm the entire resonance depth.
\end{enumerate}

\subsection{Proof-of-concept measurements}
% C:\Work\Postdoc\Investigations\Experiment noise_Measurement_10GHz_cosmic_HEMT\240115_circulator_pump_scheme\with_20dB_att_3dB_splitter
% Bifurcation
\begin{figure}[!htb]
\includegraphics[width=8.6cm]{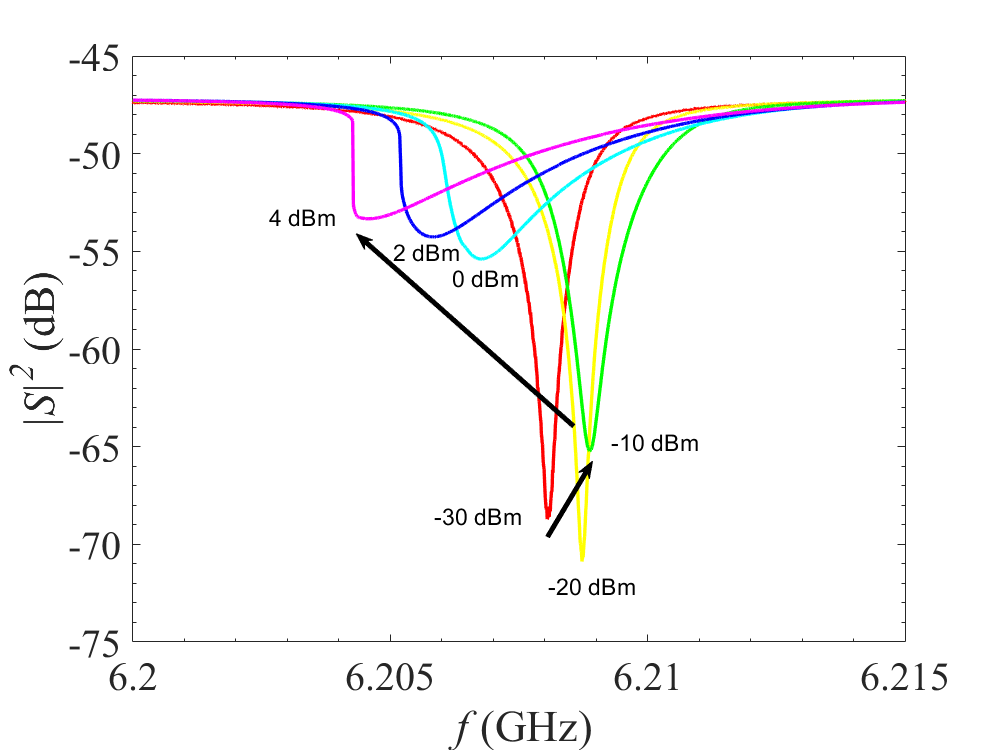}
\caption{\label{fig:2} Measurement of the resonance response of the reflection mode as a function of frequency at different sweep powers. The sweep power increases from $-30\,\mathrm{dBm}$ to $4\,\mathrm{dBm}$ as measured at the output of the vector network analyser, with the red line corresponding to the lowest power and the magenta line corresponding to the highest power. At sweep power of $2\,\mathrm{dBm}$, discontinuity starts to develop in the resonator response.}
\end{figure}

In our first set of experiments, we applied the pump separation scheme to a half-wave NbN resonator with quality factor of $Q_\mathrm{r}\sim 30000$ (when measured using a weak sweep tone). Figure~\ref{fig:2} shows measurements of S-parameter where the sweep tone was injected via port 1 of the circulator and measured via port 3 of the circulator. The ports were configured according to figure~\ref{fig:schematic_diagram}. The measured values were offset by a $20\,\mathrm{dB}$ attenuator before the input of the cryostat, coaxial cable loss in the cryostat, and circulator loss at the cold stage of the cryostat. (This is the only measurement where there is an attenuator in the pump path. The attenuator was included here to enable measurement of the resonance at very low sweep powers.) The measured S-parameter resonance sweeps display additional features compared to the idealised kinetic inductance nonlinear behaviour shown in figure~\ref{fig:0}. Firstly, the depth of the resonance decreases with increasing power. This indicates increase in dissipation with readout power, which may be explained by the non-equilibrium quasiparticle heating mechanism \citenumns{Goldie_2012,Visser_2014,Zhao_2022}. Additionally, the shift in resonance frequency was initially in the positive frequency direction and then changed back to the negative frequency direction as the power was increased further. This complicated effect has previously been observed and discussed in \citenumns{Zhao_2023} in the context of Ti, Nb, and NbN resonators. Overall, at high powers, the nonlinearity is dominated by the kinetic inductance nonlinearity and displays its key features such as the bifurcation discontinuity. As seen in figure~\ref{fig:2}, when the sweep power increases to $2\,\mathrm{dBm}$ and above, the S-parameter shows a discontinuity, which is the reactive bifurcation. 

The fact that the depth of resonance becomes shallower as the sweep power increases indicates a decrease in the internal quality factor of the resonator as a result of increased dissipation. From well-established resonator theories \citenumns{Swenson_2013,Thomas_2020}, the depth of the resonance is given by
\begin{align}
 |S|^2 &= \left(1- \frac{Q_\mathrm{r}}{Q_\mathrm{c}}\right)^2\,, \label{eq:Res}
\end{align}
where $Q_\mathrm{r}$ is the total quality factor given by $Q_\mathrm{r}^{-1}=Q_\mathrm{i}^{-1}+Q_\mathrm{c}^{-1}$, $Q_\mathrm{i}$ is the internal quality factor, and $Q_\mathrm{c}$ is the coupling quality factor. Since the pump separation scheme relies on placing the pump tone near the bottom of the resonance sweep, this shallowing of the resonance depth will place a limit on degree of pump separation.

% C:\Work\Postdoc\Investigations\Experiment noise_Measurement_10GHz_cosmic_HEMT\240115_circulator_pump_scheme\with_20dB_att_3dB_splitter
% Gain and isolation
\begin{figure}[!htb]
\includegraphics[width=8.6cm]{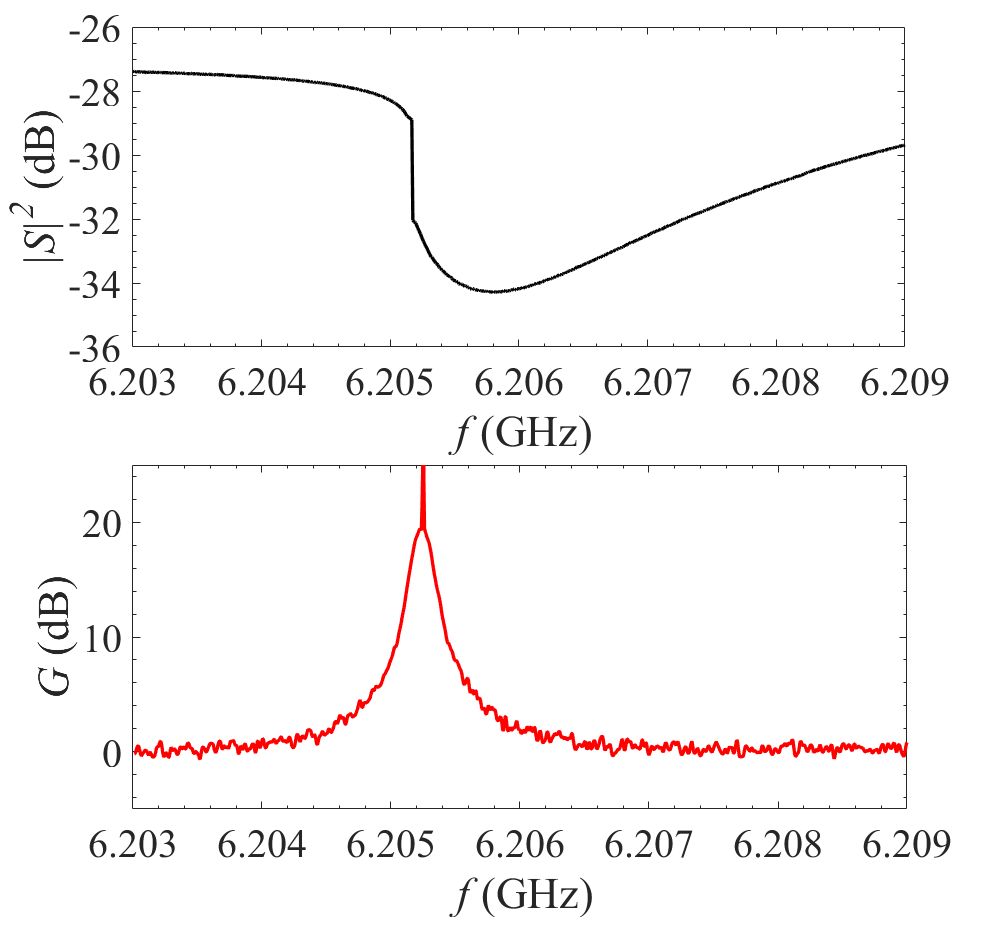}
\caption{\label{fig:3} Measurement of signal amplification and pump separation for the proof-of-concept resonator amplifier at $0.1\,\mathrm{K}$. The operating point is given by pump power of $-18.00\,\mathrm{dBm}$ measured just outside of the cryostat, and pump frequency of $6.20525\,\mathrm{GHz}$. Top subfigure: resonance response of the reflection mode where the sweep power is equal to the pump power of the operating point. Bottom subfigure: gain profile of the signal tone as a function of frequency. The pump tone is fixed in power and frequency at the operating point, and it is represented by the spike in the bottom subfigure. The peak gain of the bottom subfigure is $19\,\mathrm{dB}$. In the top subfigure, far from resonance, a reflection coefficient of $-27\,\mathrm{dB}$ is measured; at the operating point of the pump tone, a decreased reflection coefficient of $-32.5\,\mathrm{dB}$ is measured instead. Thus $5.5\,\mathrm{dB}$ of pump separation has been achieved at this operating point. }
\end{figure}

We then operated the same resonator as an amplifier, choosing the operating point using the pump separation scheme described earlier. The top subfigure of figure~\ref{fig:3} shows frequency sweep of the resonance using the pump tone as the sweeping tone. The pump tone has the same power as the amplifier operating point and is the only tone present in this resonance sweep. The bottom subfigure shows the amplification profile of the signal tone with the pump tone fixed at the operating point. The sharp spike at the centre of the bottom subfigure indicates the frequency location of the strong pump tone and does not represent actual gain. The operating point was given by using a pump power of $-18.00\,\mathrm{dBm}$ measured just outside of the cryostat, and a pump frequency of $6.20525\,\mathrm{GHz}$. Far from resonance, the full power of the pump was reflected, giving a reflection coefficient of $-27\,\mathrm{dB}$; at the operating point, most of the pump power is transmitted and a decreased reflection coefficient of $-32.5\,\mathrm{dB}$ was obtained instead. The same reflection coefficient was obtained both by reading off the resonance sweep at the frequency of the operating point (top subfigure) as well as by measuring the pump S-parameters during an amplification sweep with the pumping VNA set to continuous wave mode (bottom subfigure). Far from resonance, the resonator reflects all the incident power, and the reflection coefficient is determined mainly by coaxial cable loss in the cryostat. This value can therefore be used as a normalisation factor to determine the amount of pump power reflected from the resonator relative to its input, in this case $-32.5\,\mathrm{dB} + 27\,\mathrm{dB} = -5.5\,\mathrm{dB}$. As expected from the design of the pump separation scheme, the signal tone was amplified strongly around the pump tone with a maximum gain of $\sim19\,\mathrm{dB}$ while the reflected pump tone was reduced by $5.5\,\mathrm{dB}$ due to its placement near the bottom of the resonance in reflection. Our pump separation scheme turned out to be highly successful. Nevertheless, we note that the degree of separation is limited by the depth of the resonance, which is in turn limited by the relative magnitudes of the coupling and intrinsic quality factors as discussed above.

\subsection{Optimised Qc measurements}
% C:\Work\Postdoc\Investigations\Experiment noise_Measurement_10GHz_cosmic_HEMT\240208_QC008
% Gain and isolation
\begin{figure}[!htb]
\includegraphics[width=8.6cm]{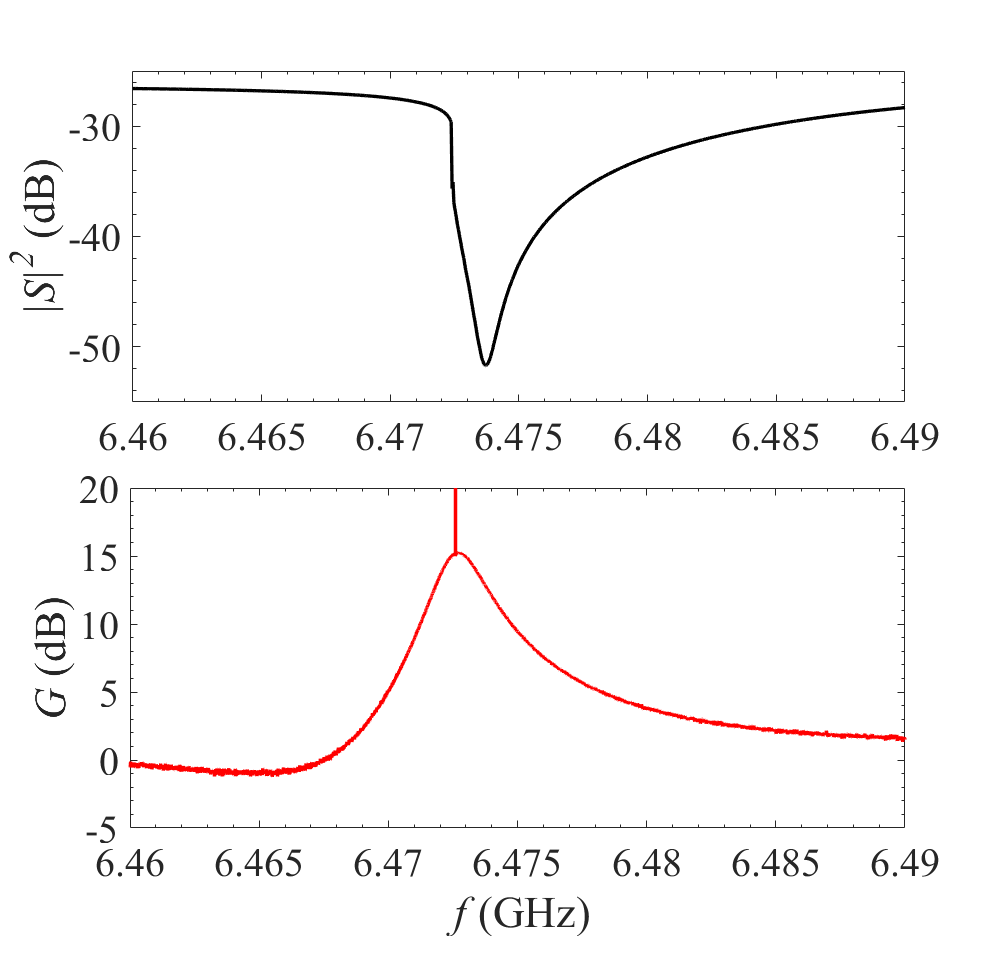}
\caption{\label{fig:4} Measurement of signal amplification and pump separation for the optimised resonator amplifier at $0.1\,\mathrm{K}$. The operating point is given by pump power of $-12.50\,\mathrm{dBm}$ measured just outside of the cryostat, and pump frequency of $6.4726\,\mathrm{GHz}$. Top subfigure: resonance response of the reflection mode where the sweep power is equal to the pump power of the operating point. Bottom subfigure: gain profile of the signal tone as a function of frequency. The pump tone is fixed in power and frequency at the operating point, and it is represented by the spike in the bottom subfigure. The peak gain of the bottom subfigure is $15\,\mathrm{dB}$. In the top subfigure, far from resonance, a reflection coefficient of $-26.5\,\mathrm{dB}$ is measured; at the operating point of the pump tone, a decreased reflection coefficient of $-38.5\,\mathrm{dB}$ is measured instead. Thus $12\,\mathrm{dB}$ of pump separation has been achieved at this operating point. }
\end{figure}

% C:\Work\Postdoc\Investigations\Experiment noise_Measurement_10GHz_cosmic_HEMT\240208_QC008\40dB_Att_isolation_6GHz
% Gain logarithmic
\begin{figure*}
\includegraphics[width=18cm]{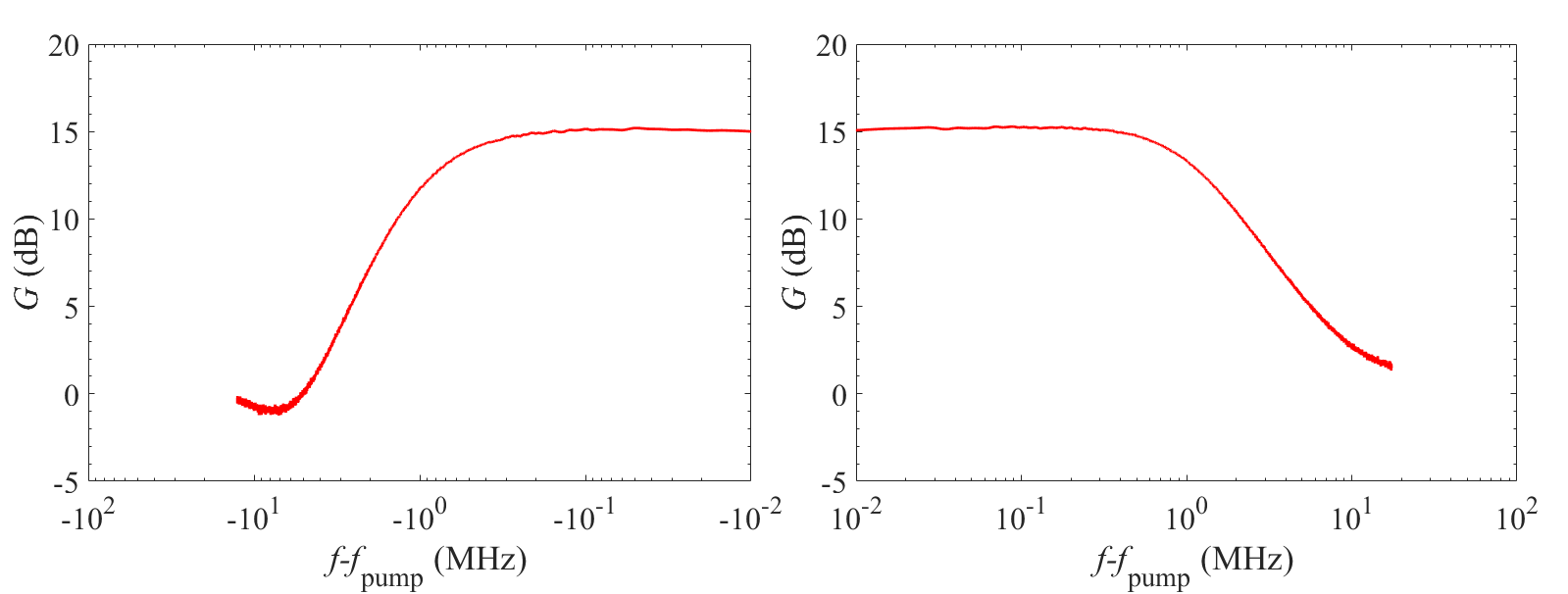}
\caption{\label{fig:5} Measurement of the gain profile of the optimised resonator amplifier at $0.1\,\mathrm{K}$. The gain of the signal tone is plotted against relative frequency $f-f_\mathrm{pump}$ on a logarithmic scale.}
\end{figure*}

The results of the previous section suggest that the pump separation scheme can be improved by increasing the depth of the resonance in reflection. With reference to equation~\ref{eq:Res}, a deep resonance can be achieved by ensuring that $Q_\mathrm{r}$ is limited by $Q_\mathrm{c}$ such that $Q_\mathrm{r}\sim Q_\mathrm{c}\ll Q_\mathrm{i}$. In other words, energy transfer out of the resonator should be dominated by capacitive coupling to the external circuit. In our design, this can be achieved by extending the lengths of the re-entrant coupling capacitors, which are shown in the bottom subfigure of figure~\ref{fig:schematic_diagram}. In addition to improving the degree of separation, reducing $Q_\mathrm{r}$ and $Q_\mathrm{c}$ has the additional advantage of increasing the bandwidth of the amplifier. This is especially important as resonator amplifiers have narrow bandwidths compared with travelling-wave amplifiers.

To verify this reasoning, we swapped the resonator of the previous section, which had a coupling length of $10\,\mathrm{\mu m}$, with a different resonator which had a coupling length of $300\,\mathrm{\mu m}$. This half-wave resonator had a quality factor of $Q_\mathrm{r}\sim 2000$ (when measured using a weak sweep tone). The top subfigure of figure~\ref{fig:4} shows frequency sweep of the resonance using the pump tone as the sweeping tone. The pump tone has the same power as the amplifier operating point and is the only tone present in this resonance sweep. The bottom subfigure shows the gain profile of the signal tone with the pump tone fixed at the operating point. The operating point was given by a pump power of $-12.50\,\mathrm{dBm}$ measured just outside of the cryostat, and pump frequency of $6.4726\,\mathrm{GHz}$. Far from resonance, the full power of the pump was reflected, giving a reflection coefficient of $-26.5\,\mathrm{dB}$ overall; at the operating point, where most of the pump tone was transmitted, a decreased reflection coefficient of $-38.5\,\mathrm{dB}$ was obtained instead. The pump tone in the reflection direction was reduced by $12\,\mathrm{dB}$ due to its placement near to the bottom of the resonance in reflection mode. In this way, we successfully increased the degree of pump separation by ensuring that $Q_\mathrm{r}$ was limited by $Q_\mathrm{c}$.

In addition to a high degree of pump separation, the optimised amplifier had a desirable signal-gain profile. Using the same data as the bottom subfigure of figure~\ref{fig:4}, figure~\ref{fig:5} shows the gain of the optimised parametric amplifier plotted against a logarithmic frequency scale. As the figure shows, the gain profile of the parametric amplifier has a flat plateau when the signal frequency is close to the pump frequency. Beyond the plateau, the gain decreases with frequency, consistent with features of a single-pole roll-off \citenumns{Zhao_2023}. The peak gain of the amplifier was $15\,\mathrm{dB}$. The $3{\text -}\mathrm{dB}$ point of the gain profile was located at $\sim1\,\mathrm{MHz}$ around the pump frequency. These characteristics are already suitable for narrow-band applications, such as direct neutrino mass measurements \citenumns{Saakyan_2020,Project8_2023_expt,QTNM_2024}. Moreover, the reproducibility of the scheme suggests that it should be possible to operate large arrays of amplifiers in a turn-key way.

% C:\Work\Postdoc\Investigations\Experiment noise_Measurement_10GHz_cosmic_HEMT\240220_QC008_Stability
% Stability
\begin{figure}[!htb]
\includegraphics[width=8.6cm]{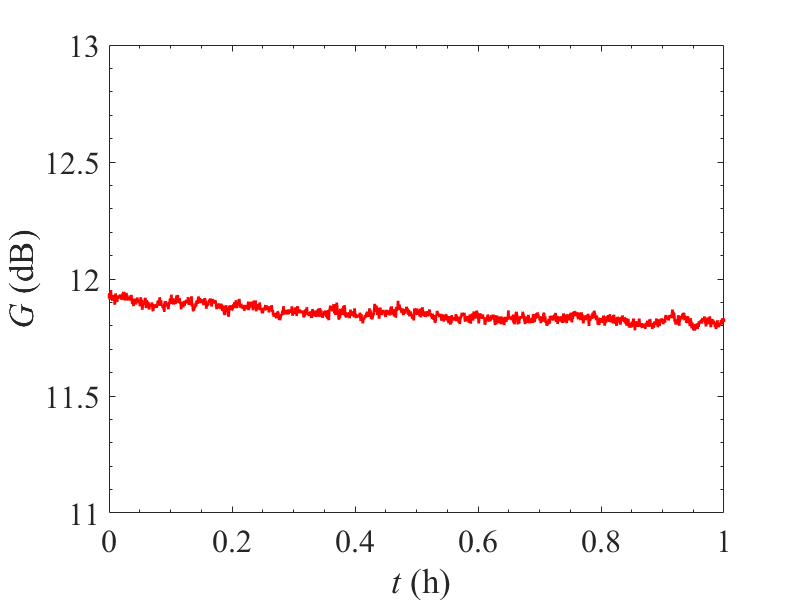}
\caption{\label{fig:6} Measurement of gain stability of the optimised resonator amplifier. The signal frequency is fixed at $6.3717\,\mathrm{GHz}$, where the gain is close to $12\,\mathrm{dB}$, and the signal gain measured continuously as a function of time for an hour. The amount of drift in gain is approximate $0.15\,\mathrm{dB}$ over an hour.}
\end{figure}

In order to investigate the characteristic stability of the amplifier over hour-long timescales, we fixed the signal frequency at $6.3717\,\mathrm{GHz}$, where the gain was close to $12\,\mathrm{dB}$, and monitored the gain continuously as a function of time. The measured drift was approximately $0.15\,\mathrm{dB}$ over a period of an hour. Based on our experience with resonator amplifiers, this remarkable stability can likely be further improved by reducing the power required to achieve bifurcation, which can be achieved by using more highly resistive films or by reducing the cross-sectional area of the central conductor of the CPW.

% C:\Work\Postdoc\Investigations\Experiment noise_Measurement_10GHz_cosmic_HEMT\240229_4K_isolation
% 4K Separation
\begin{figure}[!htb]
\includegraphics[width=8.6cm]{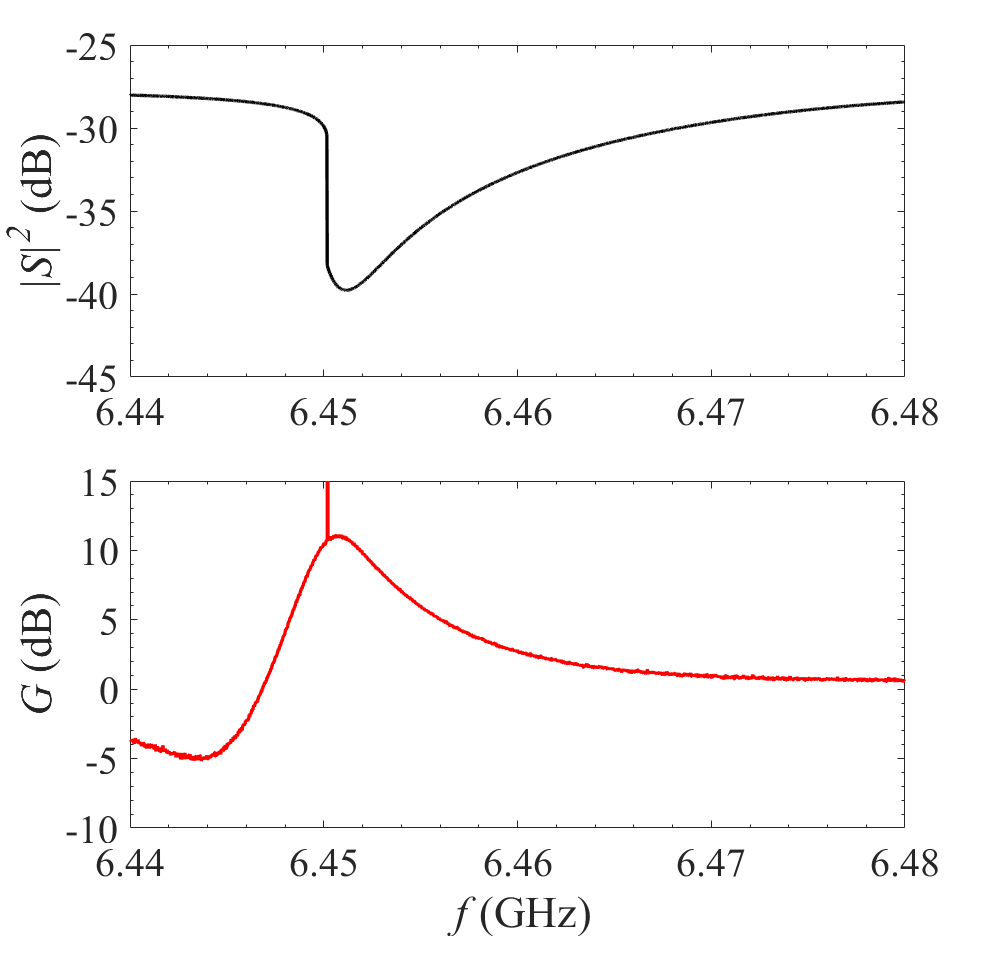}
\caption{\label{fig:7} Measurement of signal amplification and pump separation for the optimised resonator amplifier at $3.2\,\mathrm{K}$. This temperature is maintained using a pulse tube cooler. The operating point is given by pump power of $-8.50\,\mathrm{dBm}$ measured just outside of the cryostat, and pump frequency of $6.4502\,\mathrm{GHz}$. Top subfigure: resonance response of the reflection mode where the sweep power is equal to the pump power of the operating point. Bottom subfigure: gain profile of the signal tone as a function of frequency. The pump tone is fixed in power and frequency at the operating point, and it is represented by the spike in the bottom subfigure. The peak gain of the bottom subfigure is $11\,\mathrm{dB}$. In the top subfigure, far from resonance, a reflection coefficient of $-28\,\mathrm{dB}$ is measured; at the operating point of the pump tone, a decreased reflection coefficient of $-38.5\,\mathrm{dB}$ is measured instead. Thus $10.5\,\mathrm{dB}$ of pump separation has been achieved at this operating point.}
\end{figure}

Our medium-term goal is to fabricate thin-film amplifier chains that do not require HEMTs at all. We believe that this would open the door to large-format arrays that could be produced and cooled inexpensively; for example to enable the realisation of quantum-noise limited phased imaging arrays. Developing quantum limited amplifiers that reliably operate at pulse tube cooler temperatures and above is an important theme in the ongoing research of parametric amplifiers \citenumns{Malnou_4K_Paramp,Chesca_YBCO_JPA_2022}. To give an indication of whether multistage parametric amplifiers would be feasible, we operated the resonator amplifier described above at a physical temperature of $3.2\,\mathrm{K}$, which is readily achievable using a pulse tube cooler alone. The top subfigure of figure~\ref{fig:7} shows frequency sweep of the resonance using the pump tone as the sweeping tone. The pump tone has the same power as the amplifier operating point and is the only tone present in this resonance sweep. The bottom subfigure shows the amplification profile of the signal tone with the pump tone fixed at the operating point. The operating point was achieved using a pump power of $-8.50\,\mathrm{dBm}$ measured just outside of the cryostat, and a pump frequency of $6.4502\,\mathrm{GHz}$. With these operating conditions, a peak signal gain of $11\,\mathrm{dB}$ was achieved, as shown in the lower figure. Far from resonance, the full power of the pump was reflected, giving a reflection coefficient of $-28\,\mathrm{dB}$; at the operating point, most of the pump tone was transmitted, a decreased reflection coefficient of $-38.5\,\mathrm{dB}$ was obtained instead. The pump tone in the reflection direction was reduced by $10.5\,\mathrm{dB}$ due to its placement near the resonance depth of the reflection mode. Thus both high gain and a high degree of pump separation can also be achieved at temperatures maintained by a pulse tube cooler. Overall, this measurement demonstrated convincingly that a low-temperature amplifier, $0.1\,\mathrm{K}$, could also be operated easily, using the same set-up and method, at $3.2\,\mathrm{K}$.

\section{Discussion and conclusion}

We have designed, realised, and investigated a scheme for configuring superconducting resonator parametric amplifiers in a way that achieves a strong separation between the pump and signal tones. Guided by our previous theoretical work on resonator parametric amplifiers \citenumns{Thomas_2022}, our scheme exploits the intrinsic asymmetry between the S-parameters of the pump and signal tones when a half-wavelength resonator is used in reflection. By utilising the reflection mode of a superconducting two-port resonator amplifier as the output through the use of a cryogenic circulator, our scheme enables strong amplification of the signal tone and strong reduction in the power of the pump tone. Compared with our previous interference-based approach, our new circulator-based scheme does not require the delicate matching and monitoring of two, long interfering paths. The new approach has turned out to be effective, rugged, and significantly reduces complexity and cost. It eliminates a source of instability caused by phase-to-amplitude conversion in the previously used interference method. Measurements on a proof-of-concept NbN amplifier has shown that the pump separation scheme works well, but the degree of separation is limited when the device is not coupling-Q limited, and so has a shallow resonance. We predicted that the degree of pump separation could be increased by increasing the lengths of the re-entrant transmission lines used to realise the coupling capacitors in our half-wavelength resonators. Our measurements on a device with longer re-entrant coupling lines confirmed this expectation. The optimised amplifier, measured at $0.1\,\mathrm{K}$, had peak signal gain of $15\,\mathrm{dB}$ whilst achieving a pump separation of $12\,\mathrm{dB}$. The amplifier was stable during continuous operation, having a gain drift of $0.15\,\mathrm{dB}$ over an hour, with no other stabilising methods employed. The same amplifier was also operated at $3.2\,\mathrm{K}$ and achieved a peak signal gain of $11\,\mathrm{dB}$ whilst having a pump separation of $10.5\,\mathrm{dB}$. These results, at both $0.1\,\mathrm{K}$ as well as $3.2\,\mathrm{K}$, compare favourably with the previously used pump cancellation method based on destructive interference. We predict that even higher degrees of pump separation should be possible by using resonators with stronger coupling. A pump separation factor of $>10\,\mathrm{dB}$ is already a highly significant result as it is sufficient to prevent the saturation of the second-stage HEMTs in typical readout systems. The simplicity of the method, together with the demonstration of good performance at $3.2\,\mathrm{K}$, opens the door to more complicated multi-channel integrated systems for quantum sensing and computing; for example of the kind that are needed for a class of inward looking quantum-noise limited phased arrays \citenumns{withington2024quantum}. Multistage amplifiers that eliminate the need for a HEMT amplifier altogether are starting to look feasible. Future research should be carried out to build on the core idea of exploiting the asymmetry between the pump and the signal tones whilst exploring more sophisticated configurations of resonators and couplers. For example, improving the bandwidth of kinetic inductance resonator amplifiers through the use of impedance transformers, a technique previously demonstrated with great success in the context of Josephson Parametric Amplifiers \citenumns{Roy_JPA_2015,Duan_JPA_2021}, will improve the utility of these devices. Finally, measurement of the added noise from different configurations of these resonator parametric amplifiers will be important in evaluating the optimal configuration for each specific context of applications.

\begin{acknowledgements}
The authors are grateful for funding from the UK Research and Innovation (UKRI) and the Science and Technology Facilities Council (STFC) through the Quantum Technologies for Fundamental Physics (QTFP) programme (Project Reference ST/T006307/2).
\end{acknowledgements}

\bibliographystyle{h-physrev}
\bibliography{library}
\end{document}